\pdfoutput=1

\documentclass[prb,amsmath,amssymb,amsbsy,twocolumn,superscriptaddress,showpacs]{revtex4-1}

\usepackage[svgnames]{xcolor}
\definecolor{darkblue}{RGB}{0 60 120}
\definecolor{eggplant}{RGB}{190 10 150}
\usepackage{amsfonts}
\usepackage{amsmath}
\usepackage{amssymb}
\usepackage{bm}
\usepackage{epic}
\usepackage{graphicx}
\usepackage[breaklinks, colorlinks = true,linkcolor=eggplant,urlcolor=darkblue,citecolor=eggplant]{hyperref}
\usepackage{mathrsfs}
\usepackage{soul}
\usepackage[lofdepth,lotdepth,caption=false]{subfig}
\usepackage{varwidth}
\usepackage{wrapfig}
\usepackage{times}
\usepackage{bm}
\usepackage{epstopdf}
\usepackage[abs]{overpic}
\newcommand{\liiro}{$\beta$-Li${}_{2}$IrO${}_{3}$}
\newcommand{\liirotitle}{$\boldmath{\beta}\text{-Li}{}_{\boldmath{2}}\text{IrO}{}_{\boldmath{3}}$}
\newcommand{\jhalf}{$j_{\text{eff}}=1/2$}
\newcommand{\jhalftitle}{$\boldmath{j_{\text{eff}}=1/2}$}
\newcommand{\jthreehalf}{$j_{\text{eff}}=3/2$}

\begin{document}

\title{Topological and magnetic phases with strong spin-orbit coupling
  on the hyperhoneycomb lattice}

\author{Eric Kin-Ho Lee}
\affiliation{Department of Physics and Center for Quantum Materials,
University of Toronto, Toronto, Ontario M5S 1A7, Canada.}
\author{Subhro Bhattacharjee}
\affiliation{Department of Physics and Center for Quantum Materials,
University of Toronto, Toronto, Ontario M5S 1A7, Canada.}
\affiliation{Max-Planck-Institut f\"ur Physik komplexer Systeme, N\"othnitzer Str. 38, 01187
  Dresden, Germany} 
\author{Kyusung Hwang}
\affiliation{Department of Physics and Center for Quantum Materials,
University of Toronto, Toronto, Ontario M5S 1A7, Canada.}
\author{Heung-Sik Kim}
\affiliation{Department of Physics, Korean Advanced Institute of Science and Technology, Daejun 305-701, Korea}
\author{Hosub Jin}
\affiliation{Center for Correlated Electron Systems, Institute for Basic Science (IBS), Seoul 151-747, Korea}
\affiliation{Department of Physics and Astronomy, Seoul National University, Seoul 151-747, Korea}
\author{Yong Baek Kim}
\affiliation{Department of Physics and Center for Quantum Materials,
University of Toronto, Toronto, Ontario M5S 1A7, Canada.}
\affiliation{School of Physics, Korea Institute for Advanced Study, Seoul 130-722, Korea.}

\begin{abstract}
  We study the general phase diagram of correlated electrons for
  iridium-based (Ir) compounds on the hyperhoneycomb lattice---a
  crystal structure where the Ir$^{4+}$ ions form a three-dimensional
  network with three-fold coordination recently realized in the
  \liiro{} compound. Using a combination of microscopic derivations,
  symmetry analysis, and density functional calculations, we determine
  the general model for the electrons occupying the \jhalf{} orbitals
  at the Ir$^{4+}$ sites.  In the non-interacting limit, we find that
  this model allows for both topological and trivial electronic band
  insulators along with metallic states. The effect of Hubbard-type
  electron-electron repulsion on the above electronic structure in
  stabilizing $\mathbf{q}=\mathbf{0}$ magnetic order reveals a phase
  diagram with continuous phase transition between a topological band
  insulator and a N\'eel ordered magnetic insulator.
\end{abstract}
\date{\today}
\maketitle

\section{\label{sec:intro}Introduction}

The importance of the interplay between spin-orbit coupling (SOC) and
electron-electron correlations in stabilizing a wide variety of novel
electronic phases such as topological insulators (TI), Weyl
semi-metals, and quantum spin liquids has been explored
recently.\cite{PhysRevLett.102.017205,yang2010topological,pesin2010mott,PhysRevLett.105.027204,wan2011topological,witczak2013correlated,nussinov2013compass}
Materials such as 5d transition metal (iridium=Ir, osmium=Os) oxides
with strong atomic SOC provide fertile grounds to uncover the above
physics and a large number of such compounds are currently being
investigated.\cite{JPSJ.70.2880,nakatsuji2006metallic,matsuhira2007metal,kim2008novel,Kim06032009,PhysRevB.82.064412,PhysRevB.83.220403,PhysRevLett.108.127204,qi2012strong}

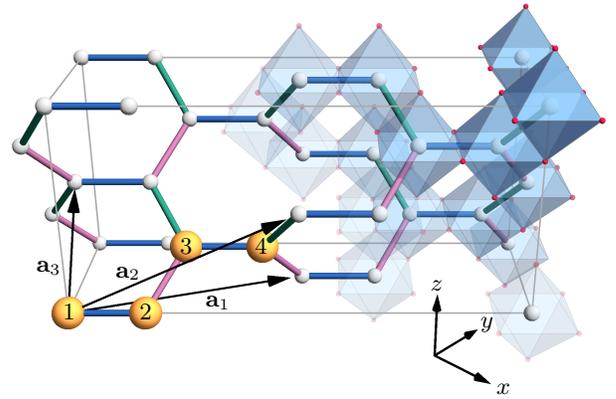
\begin{figure}[h!]
  \centering
  \setlength\fboxsep{0pt}
  \setlength\fboxrule{0pt}
  \fbox{\begin{overpic}[scale=.12,clip=true,trim=150 550 265 0]{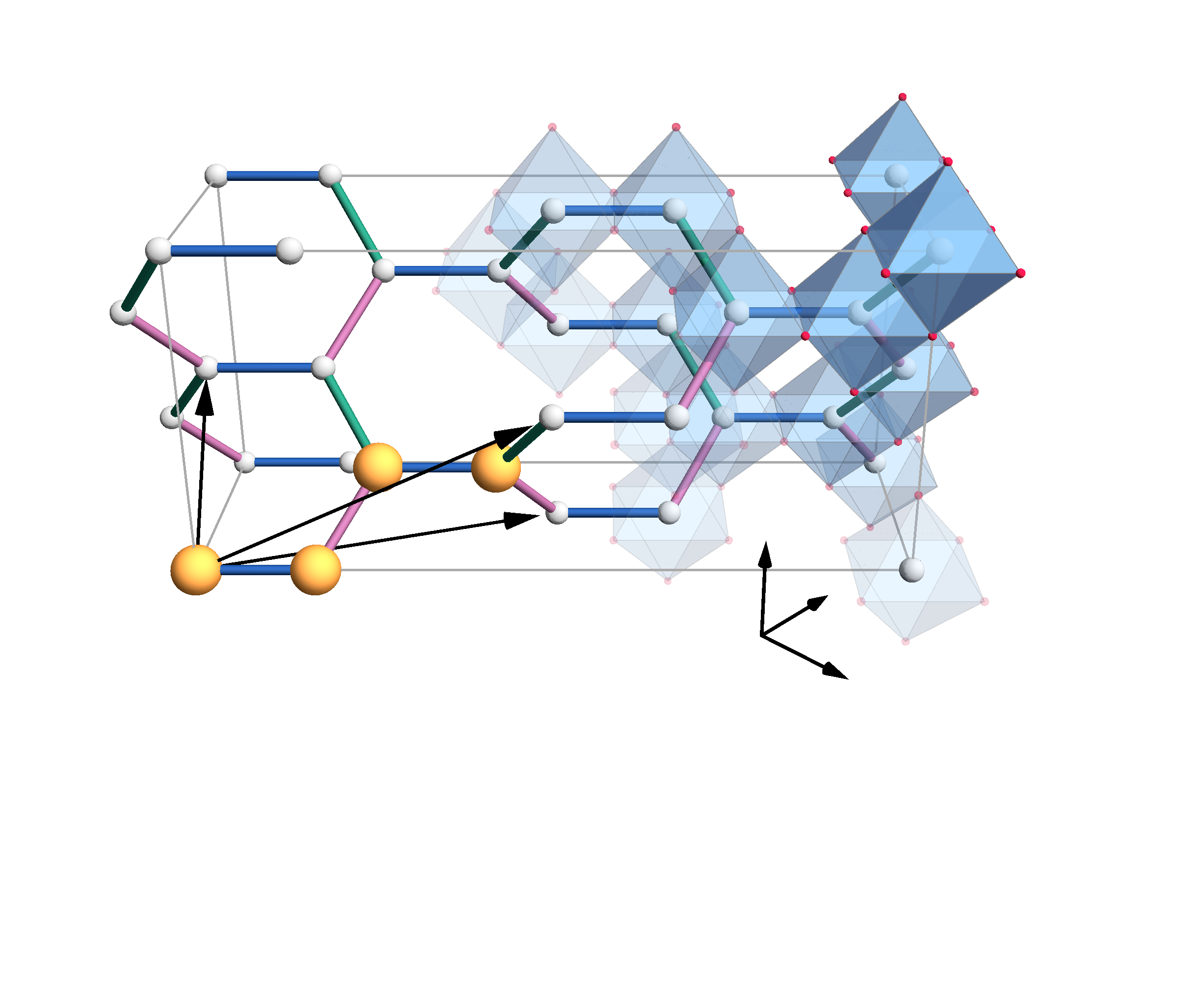}
      \put(18,53){$\mathbf{a}_3$}
      \put(82,40){$\mathbf{a}_1$}
      \put(48,51){$\mathbf{a}_2$}
      \put(192,7){$x$}
      \put(186,32){$y$}
      \put(167,46){$z$}
      \put(28,35){$1$}
      \put(57,35){$2$}
      \put(72.3,60){$3$}
      \put(100.7,60){$4$}
    \end{overpic}}
  \caption{(Color online) The ideal hyperhoneycomb lattice. The
    Ir$^{4+}$ atoms (denoted by white spheres, except for the four
    yellow ones that indicate the four atoms in our unit cell) sit in
    an octahedral cage (shaded in blue) of oxygen atoms (small red
    spheres). The lattice vectors are denoted by ${\bf a}_1, {\bf
      a}_2$ and ${\bf a}_3$.  The three nearest-neighbor bonds are
    referred to as $x$ (green), $y$ (pink) and $z$ (blue) bonds.}
  \label{fig:lattice}
\end{figure}

Recently, the material $\beta$-Li$_2$IrO$_3$ has been synthesized by
Takagi {\it et al.}\cite{2013_takagi} which has attracted attention
due to the novel three-dimensional network formed by the Ir$^{4+}$
ions---the hyperhoneycomb lattice (see Fig. \ref{fig:lattice}).  It
has been theoretically predicted that the spin model in the
strong-coupling limit can be highly anisotropic and may lead to
interesting magnetic as well as a three-dimensional Kitaev quantum
spin-liquid ground
state.\cite{PhysRevB.79.024426,PhysRevB.89.045117,PhysRevB.89.014424,kimchi2013}

In this paper, motivated by the above developments, we study the weak-
and intermediate-coupling regimes of \liiro{} and iso-structural
compounds with Ir situated on a hyperhoneycomb lattice.  We point out
the possibility of interesting ground states in these systems that
generally arise from the nature of the underlying lattice geometry and
strong SOC effects.  In turn, these results can shed light on the
physics of the above material and others on a similar lattice
structure.

An important starting point in the study of these compounds is to
ascertain the nature of the electronic structure, particularly that of
the electronic bands near the Fermi level. Due to the large atomic
SOC, as in a large number of Ir-based
compounds,\cite{Kim06032009,PhysRevB.82.064412,pesin2010mott,witczak2012topological,witczak2013correlated}
the low energy bands are expected to be formed by \jhalf{} atomic
orbitals. Using the symmetries of the hyperhoneycomb lattice, we
obtain the general tight-binding Hamiltonian for the \jhalf{}
orbitals. Apart from the generic metal and band insulator (BI), we
find that this hopping Hamiltonian allows for a three-dimensional
strong TI (STI) over a large parameter regime. The above tight-binding
model is further justified by more microscopic calculations based on
Slater-Koster parameters for the $5d$ orbitals in the large SOC limit
for the \textit{ideal} hyperhoneycomb lattice. This latter calculation
also reveals the connection between the symmetry-allowed hopping
parameters and the Slater-Koster parameters. In parallel, we perform
density functional theory (DFT) calculations in the presence of SOC to
probe the nature of the states near the Fermi level for \liiro{} on an
ideal hyperhoneycomb lattice. The DFT results support our assumption
that the low energy states near the Fermi level have a predominantly
\jhalf{} orbital character and are well separated from the
\jthreehalf{} bands that lie below the Fermi level. The study of the
$\mathbf{q}=\mathbf{0}$ magnetic phases induced by Hubbard-type
electronic correlations on the above electronic structure reveals an
interesting phase diagram. We find a direct continuous transition
between the STI at weak correlations and magnetic insulator with
N\'eel order at intermediate correlations.  Although the metallic
state in the weak-coupling limit ultimately transitions into the
N\'eel ordered magnetic insulator at sufficiently large correlations,
an intermediate phase---a magnetically-ordered (N\'eel) metal---is
first reached via a discontinuous transition.  Interestingly, while
time-reversal and inversion symmetries are broken in the magnetically
ordered insulator, the product of the two is found to be preserved,
leading to pseudo-Kramers doublets in the energy spectrum.

The rest of the paper is organized as follows. We begin with a
discussion of the ideal hyperhoneycomb lattice and its symmetries in
Sec. \ref{sec:lattice}. Using these symmetries, the general tight
binding model (up to second-nearest-neighbor, 2NN) for the \jhalf{}
orbitals is then obtained in Sec. \ref{sec:tb}. The hopping
Hamiltonian contains both spin conserving (scalar) as well as
spin-flipping (vector) hopping amplitudes. While we show that the
nearest-neighbor (1NN) vector hopping terms are inconsequential, the
2NN hopping terms can stabilize a three-dimensional STI over a large
parameter regime. We study the detailed phase diagram of the symmetry
allowed tight-binding Hamiltonian in Sec. \ref{sec:2NNpd} and point
out a simple relation in the hopping parameters that separates the
trivial and the topological band insulators in the phase diagram. In
Sec. \ref{sec:micro}, we establish the connection between the symmetry
allowed hopping parameters and the more microscopic Slater-Koster
parameters characterizing the hopping Hamiltonian for the underlying
$t_{2g}$ bands. In this section, starting from such a hopping
Hamiltonian on an ideal hyperhoneycomb lattice and taking the large
SOC limit, we derive the pertinent Hamiltonian for the \jhalf{}
orbitals to the leading order of perturbation theory. In
Sec. \ref{sec:fit}, we show the results of fully relativistic DFT
calculations on \liiro{} assuming an ideal hyperhoneycomb lattice for
the material. These calculations reveal the generic separation of the
\jhalf{} bands and the \jthreehalf{} bands with the former being
closer to the Fermi level, justifying our generic \jhalf{}
tight-binding calculations in earlier sections. Further, fitting the
DFT band structure with the $t_{2g}$ tight-binding model, we obtain an
estimate of the parameter regime of the tight-binding model which may
be relevant to \liiro. After completing the characterization of the
low energy electronic structure, in Sec. \ref{sec:mag}, we study the
effect of short range electron-electron interactions in the
intermediate correlation regime.

\section{\label{sec:lattice}The ideal hyperhoneycomb lattice of ${\rm \bf
    Ir}^{\boldmath{4+\text{}}}$ ions}

We first consider the generic symmetry-allowed \jhalf{} tight-binding
model for the network of Ir in the hyperhoneycomb lattice. As
described below and also supported by our DFT calculations in
Sec. \ref{sec:fit}, these \jhalf{} orbitals are expected to form the
low energy electronic excitations near the Fermi level. To this end,
we start with a description of the hyperhoneycomb lattice and its
symmetries.

The hyperhoneycomb lattice consists of a network of Ir$^{4+}$ ions
where each Ir$^{4+}$ ion has three 1NNs and sits in an octahedral oxygen cage
(Fig. \ref{fig:lattice}). A detailed structural description of the
lattice can be obtained from the x-ray diffraction experiments on
$\beta$-Na$_2$PtO$_3$\cite{urland1972kenntnis} which belongs to the
same space-group ($Fddd$) as \liiro{}.  In the ideal structure, which
we refer in this paper as the \textit{ideal} hyperhoneycomb lattice
(shown in Fig. \ref{fig:lattice}), the oxygen octahedra are
undistorted and the Ir-O-Ir and the Ir-Ir-Ir bond angles measure
90$^\circ$ and 120$^\circ$ respectively and hence different Ir-Ir
bonds have same length.

The lattice structure can be described as a face-centred orthorhombic
lattice with four Ir sites per unit cell.\cite{PhysRevB.89.045117}
While a complete discussion of the symmetries of the lattice is given
in Ref. \onlinecite{PhysRevB.89.045117}, here we note that since the
hyperhoneycomb lattice possesses inversion symmetry, the eight bands
arising from two \jhalf{} orbitals at each of the four sublattices
become four doubly degenerate bands due to Kramers theorem. In the
following, we also exploit the presence of this inversion symmetry by
using parity eigenvalues when computing the Z$_2$ topological
invariants.\cite{fu2007topological} Out of the three 1NN bonds (which
we call the $x,y$ and $z$, following notation used in the
Heisenberg-Kitaev model explored in
Ref. \onlinecite{PhysRevB.89.045117} and
Ref. \onlinecite{PhysRevB.89.014424}, see fig. \ref{fig:lattice}), two
of the bonds (namely $x$ and $y$) are equivalent due to the presence
of C$_2$ symmetry.  More details regarding the lattice used in our
ideal hyperhoneycomb calculations can be found in Appendix
\ref{app:lattice}.

At each Ir$^{4+}$ site, the octahedral crystal field of the oxygen
splits the $5d$ Ir orbitals into the upper $e_g$ orbitals (4-fold
degenerate including spin degeneracy) and the low lying $t_{2g}$
orbitals (6-fold degenerate including spin degeneracy) with the
separation ($10Dq$) being approximately $3~eV$. Neglecting the
$t_{2g}-e_{g}$ mixing due to large energy separation, the strong SOC
($\lambda\sim 500~meV$), splits the six $t_{2g}$ orbitals into the low
energy four \jthreehalf{} quadruplet and high energy \jhalf{}
doublet. The five $5d$ electrons completely fill up the quadruplet
leaving the doublet half-filled.  These half-filled \jhalf{} atomic
orbitals, one at each Ir$^{4+}$ site, form the low energy electronic
degrees of freedom in this compound.

A few remarks are in order before we proceed to the description of the
tight-binding and DFT results. While the symmetry-allowed
tight-binding model described in Sec. \ref{sec:tb} is generally valid
for changes in the position of both oxygen and the Ir$^{4+}$ ions as
long as the space group ($Fddd$) remains intact and the \jhalf{} bands
remain well-separated from the \jthreehalf{} bands, the microscopic
calculations starting from the $t_{2g}$ orbitals presented in
Sec. \ref{sec:micro} assume ideal position of the oxygen atoms which
in turn affect the overlap integrals. The effective hopping
Hamiltonian for the \jhalf{} derived from it assumes that the the
leading order corrections due to SOC coupling effects are captured
within a second order perturbation theory which is valid in the large
SOC limit. Since the detailed structure of \liiro{} is not available
at present and also future compounds may differ by small details in
the structure such as the position of oxygen ions, we start with the
most general case in Sec. \ref{sec:tb} and specialize to the ideal
hyperhoneycomb lattice later.  Our DFT calculations in
Sec. \ref{sec:fit}, based on the ideal structure for \liiro{},
validates our above assumption of the separation of the \jhalf{} and
\jthreehalf{} bands in that limit.
\section{\label{sec:tb}Symmetry-allowed tight-binding model}

Using various symmetries of the lattice discussed above, we can write
down the generic tight-binding model for the \jhalf{} electrons on the
hyperhoneycomb lattice. The general hopping Hamiltonian is given by:
\begin{align}
H_{\text{tb}}=\sum_{ij}c^{\dagger}_{i}
  h_{ij} c_{j}
  \label{eq:tbgen}
\end{align}
with
\begin{align}
h_{ij}=t_{ij}\mathbb{I}+i{\bf v}_{ij}\cdot\mathbb{\sigma}
\label{eq_hopmat}
\end{align}
where $c^{\dagger}_i=(c^{\dagger}_{i \uparrow},c^{\dagger}_{i
  \downarrow})$ are the creation operators in the \jhalf{} basis at
site $i$, $\vec{\sigma}=(\sigma_x,\sigma_y,\sigma_z)$ are the Pauli
matrices, and $\mathbb{I}$ is the $2\times2$ identity matrix. $t_{ij}$
and ${\bf v}_{ij}$ denote the scalar and the spin-flip hopping
respectively.\cite{PhysRevB.87.214416}

Below, we determine the hopping Hamiltonian up to second
nearest-neighbor (2NN). A more microscopic approach based on
Slater-Koster parameters including various hopping paths and its
connection to the symmetry-allowed hopping parameters is presented in
Sec.
\ref{sec:micro}.

\subsection{\label{subsec:tb1NN}Nearest-neighbor}

At the nearest-neighbor level (1NN), as noted earlier, there are two
symmetry-inequivalent sets of bonds to consider: the $x$/$y$-bonds and
the $z$-bonds .  The symmetry-allowed 1NN tight-binding hopping matrix
Eq. (\ref{eq_hopmat}) can be written as
\begin{align}
h_{x/y}^{\text{1NN}}= t_{xy}\mathbb{I}
\label{eq:tb1nnxy}
\end{align}
for the $x/y$ bonds and
\begin{align}
h_z^{\text{1NN}}=t_z^{\text{1NN}}\mathbb{I}+i\mathbf{v}_{z}^{\text{1NN}}\cdot \mathbb{\sigma}
\label{eq:tb1nnz}
\end{align}
for the $z$ bonds. The absence of spin-dependent vector hopping
amplitudes on the $x$- and $y$-bonds is due to inversion symmetry at
their bond centers. Each $z$-bond, on the other hand, has three $C_2$
axes passing through its bond center, which constrains the
spin-dependent vector hopping amplitudes to point in the
$\pm\left(\hat{x}+\hat{y}\right)$ direction.  We use the convention
$\hat{v}_{z}^{(12)}=(1,1,0)/\sqrt{2}$ and, by symmetry,
$\hat{v}_{z}^{(34)}=-(1,1,0)/\sqrt{2}$, where the superscripts
indicate sublattice indices that are involved in the particular
$z$-bond.

We note that in the purely 1NN model, the spin-dependent vector
hopping amplitude on the $z$-bonds can be eliminated by a
sublattice-dependent basis transformation.  To see this, we re-write
the $z$-bond hopping amplitudes Eq. (\ref{eq:tb1nnz}) as
\begin{align}
h_z^{\text{1NN}}=\sqrt{\left(t_{z}^{\text{1NN}}\right)^2+\left|{\bf v}_z^{\text{1NN}}\right|^2} e^{i \theta
    \hat{v}_{ij}\cdot \vec{\sigma} }.
\end{align}
where, $\tan\theta=|{\bf
  v}_{z}^{\text{1NN}}|/t_{z}^{\text{1NN}}$. Rotating the \jhalf{}
electrons on sublattices 2 and 3, for example, by $\exp\left(-i \theta
  \hat{v}_{z}\cdot \vec{\sigma}\right)$ would render hopping
amplitudes on the $z$-bonds diagonal in \jhalf{} pseudo-spin space
without affecting the form of the hopping on the $x$- and $y$-bonds
which are already diagonal in the pseudo-spin indices.

\begin{figure}
  \centering
  \setlength\fboxsep{0pt}
  \setlength\fboxrule{0.0pt}
  \fbox{\begin{overpic}[scale=1]{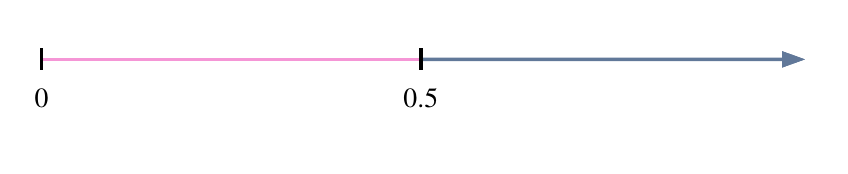}
      \put(223,20){$\infty$}
      \put(40,40){band insulator}
      \put(165,40){metal}
      \put(110,7){$t_{xy}/t_{z}$}
    \end{overpic}}
  \caption{(Color online) Phase diagram for the strictly
    nearest-neighbor hopping Hamiltonian. The band insulator is
    topologically trivial and the metal has a closed one-dimensional
    line-node forming the Fermi surface.}
  \label{fig:1NNpd}
\end{figure}
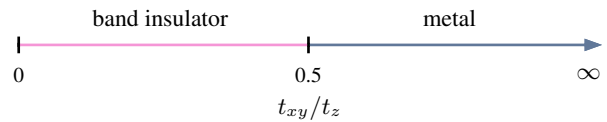

In other words, the generic symmetry-allowed 1NN \jhalf{}
tight-binding model on the hyperhoneycomb can always be written in a
SU$(2)$-invariant form (in pseudo-spin space) with the appropriate
choice of basis. Immediately, we conclude that the generic band
structure is particle-hole sym\-metric because the model is bipartite
(see end of Sec. \ref{subsec:tb2NN} and Appendix \ref{app:ph} for
general discussion on particle-hole symmetry).  In addition, all band
insulators obtained from this 1NN model would be topologically trivial
and a topologically non-trivial band insulator {\it cannot} be
realized with 1NN bonds alone. This is shown in the phase diagram
(Fig. \ref{fig:1NNpd}) of the 1NN tight-binding model at half-filling
as a function of $t_{xy}/t_{z}$.\footnote{The relative sign between
  $t_{xy}$ and $t_{z}$ can be eliminated by transforming
  $c^{\dagger}_i\rightarrow-c^{\dagger}_{i}$ for $i\in (2,3)$ while
  the overall sign is inconsequential since the Hamiltonian is
  particle-hole symmetric.} The phase diagram contains a trivial band
insulator and a metal. In the limit where $t_{xy}=0$, the
hyperhoneycomb lattice reduces to independent dimers which is a
topologically trivial insulating state with flat bands.  For
$0<2t_{xy}<t_{z}$, the flat bands disperse but the band structure
remains gapped.  At $2t_{xy}=t_{z}$, band-touching occurs at the
$\Gamma$-point.  The dispersion along the $\Gamma$-$Z$ direction is
linear near the band-touching, while it is quadratic in the
$\Gamma$-$X$ and $\Gamma$-$Y$ directions.  As $t_{xy}$ increases such
that $2t_{xy}>t_{z}$, the band-touching moves away from the
$\Gamma$-point and the Fermi surface becomes a closed line-node in the
$\Gamma$-$X$-$A_1$-$Y$-plane of the Brillouin zone. This is an
interesting feature of the strictly 1NN model that the metal has a
one-dimensional Fermi surface, {\it i.e.} a closed Fermi line-node,
instead of a regular two-dimensional Fermi surface. However, this is
not protected by symmetries and we shall find this line-node is
generally destroyed by further neighbor hopping terms.

As described in Sec. \ref{sec:micro}, starting with the $t_{2g}$
hopping Hamiltonian on the ideal hyperhoneycomb lattice and taking the
strong SOC limit, we find that, to the lowest order, the resulting
effective \jhalf{} model is described by $t_{xy}=t_{z}$ and
$\theta=0$. This falls in the metallic regime in the phase diagram
shown in Fig. \ref{fig:1NNpd}.  Here we point out that when
distortions are accounted for and/or higher order terms are included
in the perturbative series, $t_{xy}$ and $t_z$ would be in general
different and $\theta$ may be finite.
\subsection{\label{subsec:tb2NN}Second-nearest-neighbor}

For the 2NN hopping, both the scalar and spin-dependent hopping terms
are generally non-zero. There are {\it twenty} 2NN bonds in the
hyperhoneycomb lattice when using the primitive unit cell (see
Appendix \ref{app:2NN} for details). These twenty 2NN bonds can be
divided into two classes if we consider the Ir$^{4+}$ network: (1) the
2NN sites which can be connected through {\it only one} common
intermediate Ir$^{4+}$ site, and (2) the 2NN sites that requires
traversing through {\it more than one} intermediate Ir$^{4+}$ sites.
In Sec. \ref{sec:micro}, our microscopic derivation shows that,
starting from a $t_{2g}$ hopping model with 1NN hopping terms and
taking the strong SOC limit, only the twelve 2NN bonds belonging to
the first class are non-zero in the effective \jhalf{} tight-binding
model to lowest order. Hence, we shall only consider non-zero 2NN
hopping for these twelve 2NN bonds and neglect the rest in our
tight-binding model. Generally, we can write
\begin{align}
h_{ij}^{\text{2NN}}=t_{\rm 2NN}\mathbb{I}+i\left({\bf v}^{(1)}_{ij}+{\bf v}^{(2)}_{ij}\right)\cdot\mathbb{\sigma}
\label{eq:tb2nn}
\end{align}
where, $t_{\rm 2NN}$ is the scalar hopping and to bring out the
analogy with the Kane-Mele model\cite{kane2005quantum} on the
two-dimensional honeycomb lattice, we have split the spin-dependent
hopping into two parts. The first part is the three-dimensional
version of the Kane-Mele term
\begin{align}
\label{eq:km}
{\bf v}^{(1)}_{ij}=v_{\text{KM}} \frac{ \hat{r}_{ik} \times \hat{r}_{kj}}{\left| \hat{r}_{ik} \times \hat{r}_{kj} \right|}
\end{align}
where $v_{\text{KM}}$ is the strength of the coupling and ${\bf
  r}_{ik}$ and ${\bf r_{jk}=(-{\bf r}_{kj})}$ denote the vectors from
the sites $i$ and $j$ respectively to their common nearest-neighbor
site $k$. The second part of the vector hopping, not present on the
2D-honeycomb lattice (due to the presence of a mirror symmetry), is
normal to the first and is given by
\begin{align}
\label{eq:par}
{\bf v}_{ij}^{(2)}=v_{\parallel} \epsilon_{ij}  \hat{\bf r}_{ij}
\end{align}
where $v_{\parallel}$ is the strength of this coupling and
$\epsilon_{ij}=\pm 1$ is appropriately chosen such that ${\bf
  v}^{(2)}_{ij}$ transforms as a pseudo-vector under lattice
transformations as required by symmetry.

To conclude this section, we make a brief note on particle-hole
symmetry of various limits.  While the 1NN-only model is particle-hole
symmetric as mentioned in Sec. \ref{subsec:tb1NN}, finite
$t_{2\text{NN}}$ and/or $v_{\parallel}$ hopping amplitudes will break
such symmetry.  On the other hand, the $v_{\text{KM}}$ hopping
amplitude preserves this symmetry, as we show explicitly in Appendix
\ref{app:ph}.  In other words, the 1NN plus finite $v_{\text{KM}}$
model is particle-hole symmetric.

\section{\label{sec:2NNpd}Phase diagram of the 2NN tight-binding
  model}

\begin{figure}[h!]
  \centering
  \setlength\fboxsep{0pt}
  \setlength\fboxrule{0.0pt}
  \fbox{\begin{overpic}[scale=1,trim=-10 0 10 0,clip]{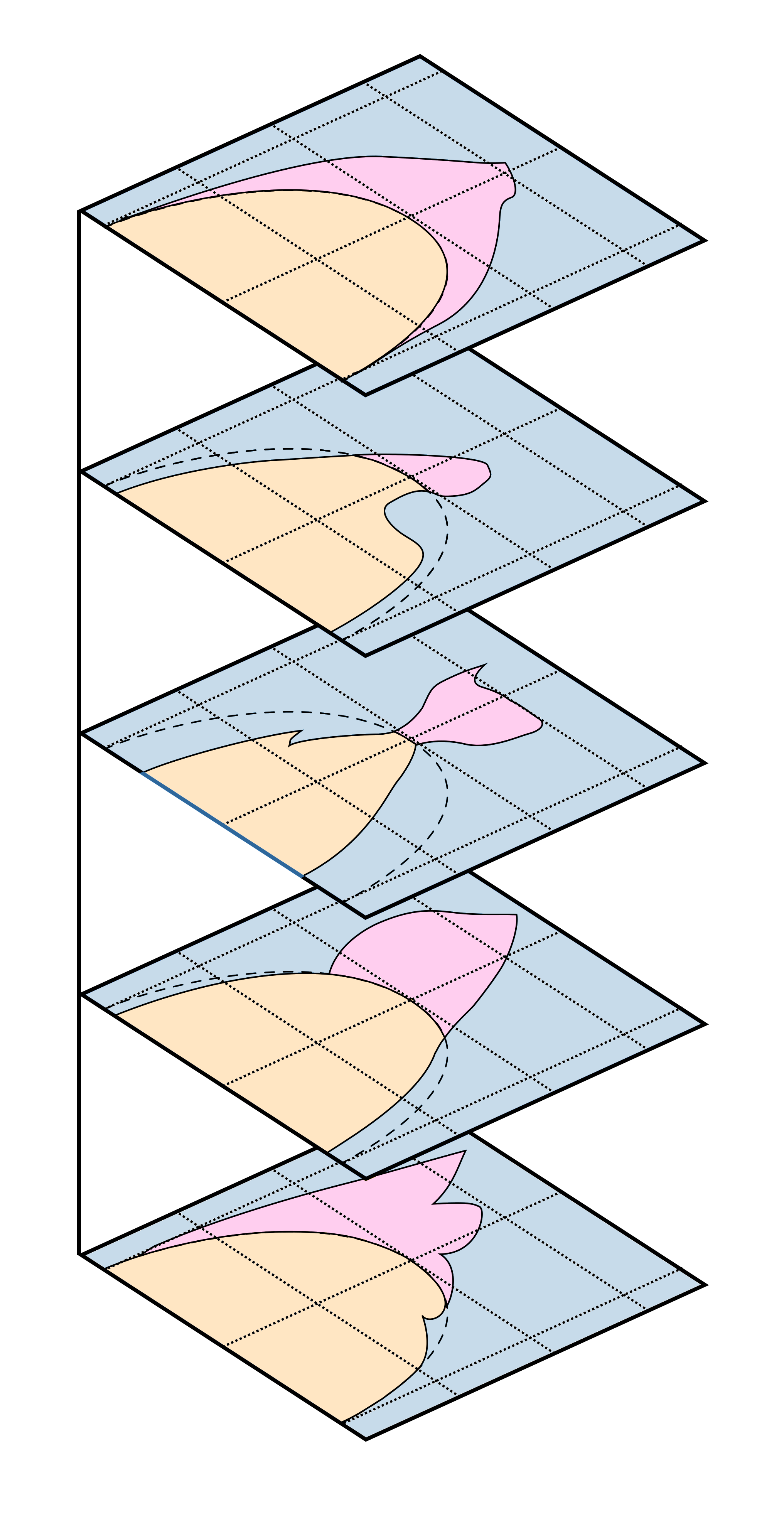}
      \put(28,70){-0.5}
      \put(66,46){0.0}
      \put(104,22){0.5}
      \put(48,37){$t_{2\text{NN}}$}
      \put(130,16){0.00}
      \put(159,30){0.25}
      \put(188,43){0.50}
      \put(218,56){0.75}
      \put(208,32){$v_{\parallel}$}
      \put(15,81){-1.0}
      \put(15,163){-0.5}
      \put(17,244){0.0}
      \put(17,326){0.5}
      \put(17,407){1.0}
      \put(0,244){$v_{\text{KM}}$}
    \end{overpic}}
  \caption{(Color online) Phase diagram in the non-interacting limit
    with 2NN hopping amplitudes with slices in the $v_{\text{KM}}$
    direction.  Nearest-neighbor hopping amplitudes have been set to
    $t_{xy}=t_{z}=1$, $v_{z}=0$.  Orange is a strong topological
    insulator, blue is a metal, and pink is a trivial band insulator.
    The dotted semi-circle indicates the region in which any
    insulating state must be a STI, and outside of which any
    insulating state must be trivial (see main text for explanation).
    The blue line along $v_{\text{KM}}=v_{\parallel}=0$ is a metallic
    state with a line-node Fermi surface akin to the metallic state in
    the 1NN-only model.}
  \label{fig:2NNpd}
\end{figure}
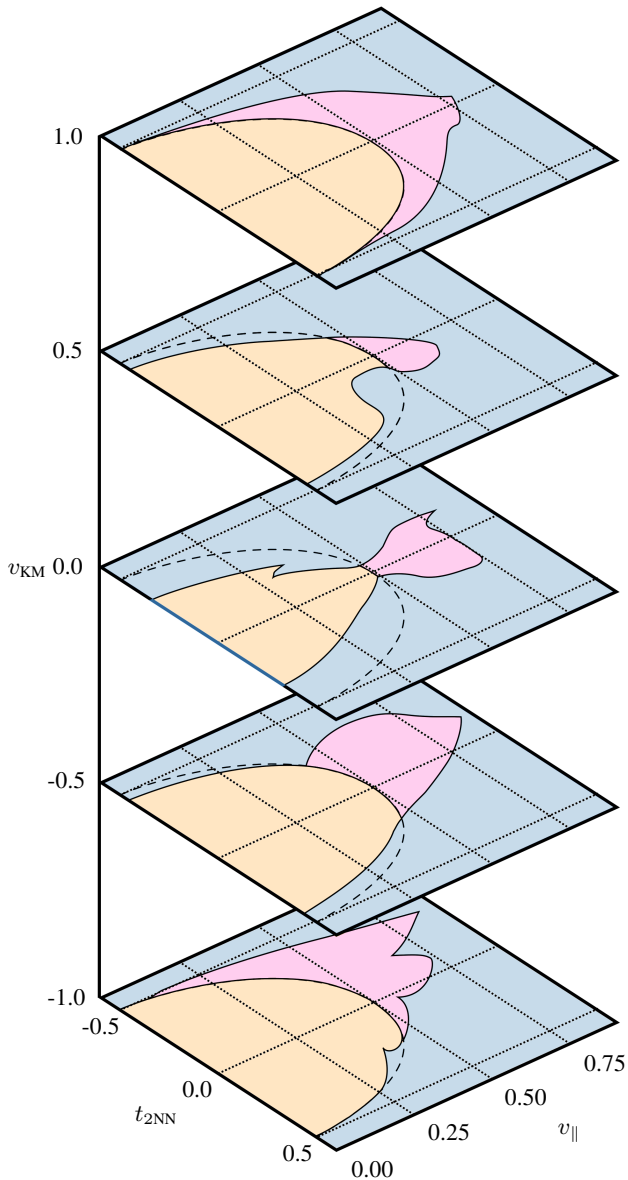

In this section, we outline the generic phase diagram for the
single-particle hopping Hamiltonian given by Eq. (\ref{eq:tbgen}), where
the different parameters are defined by Eqs. (\ref{eq:tb1nnxy}),
(\ref{eq:tb1nnz}), and (\ref{eq:tb2nn}). In Fig. \ref{fig:2NNpd}, we
present 2NN phase diagram with $t_{2\text{NN}}$, $v_{\parallel}$, and
$v_{\text{KM}}$ as the axes.  We have set the 1NN hopping integrals to
$t_{xy}=t_{z}=1$ and $\theta=0$.  We note that the phase diagram is
symmetric under
$\left(t_{2\text{NN}},v_{\text{KM}},v_{\parallel}\right)\rightarrow
-\left(t_{2\text{NN}},v_{\text{KM}},v_{\parallel}\right)$ since this
transformation merely inverts the electronic band structure (not
shown).  Hence, only $v_{\parallel}>0$ is presented.

The orange regions indicate a strong topological insulator (STI) with
Z$_{2}$ indices $(1;000)$, the blue regions indicate a metallic state,
and the pink regions indicate a trivial band insulator (BI).  The
borders between STIs and BIs are semi-metals.  This is because
time-reversal symmetry remains intact in both the phases therefore the
electronic band-gap in the bulk must close when the topology of the
bands, as encapsulated by the $Z_2$ indices, changes.  We note that
finite $t_{\text{2NN}}$ and/or $v_{\parallel}$ breaks the
particle-hole symmetry of the band structure and hence the metallic
states are generically present as opposed to the 1NN case.  In the
special case where $|t_{2\text{NN}}|$ is small and
$v_{\text{KM}}=v_{\parallel}=0$, the ground state is a metallic phase
with a line-node Fermi surface akin to the metallic phase found in the
1NN-only model.  As $|t_{2\text{NN}}|$ increases while
$v_{\text{KM}}=v_{\parallel}=0$, bands approach and cross the
Fermi level, thus generating particle and hole pockets.  This
displaces the line-node away from the Fermi level and yields a
metallic state with particle pockets.

We draw attention to the region within
$t_{2\text{NN}}^2+v_{\parallel}^2\leq 0.5^2$ (in units of $t_{xy}$;
indicated by dashed lines in Fig. \ref{fig:2NNpd}), where the STI and
metallic phases exist but not the BI phase.  On the other hand, the BI
and metallic phases can be found outside this region but not the STI
phase.  To understand this, we note that the strong $Z_2$ index is
calculated from the product of the parity
eigenvalues\cite{fu2007topological} at the time-reversal invariant
momentum (TRIM) points of the Brillouin zone (BZ) since inversion
symmetry is present. Out of the eight TRIM points in the
three-dimensional BZ, we find that the product of the parity
eigenvalues changes only at the $\Gamma$ point as we move from a STI
to a trivial band insulator. Thus we expect that the mass inversion
affects only the product of the parity eigenvalues at the $\Gamma$
point. We find that the parameter controlling this band inversion, and
hence the parity eigenvalues, depends only on $t_{2\text{NN}}$ and
$v_{\parallel}$ but not on $v_{\text{KM}}$. In fact, when $t_{\rm
  2NN}=v_\parallel=0$ and $v_{\rm KM}\neq 0$, the insulating phase is
always a STI. The accidental degeneracy that closes the band gap at
the $\Gamma$-point occurs precisely when
$t_{2\text{NN}}^2+v_{\parallel}^2 = 0.5^2$; a gap opens if we deviate
from this curve. Therefore, any insulating phase within
$t_{2\text{NN}}^2+v_{\parallel}^2\leq 0.5^2$ must have the same strong
$Z_2$ topological index as the case of $t_{\rm 2NN}=v_\parallel=0$ and
$v_{\rm KM}\neq 0$, {\it i.e.} a STI, whereas any insulating phase
outside of this region {\it can} be topologically distinct, as in this
case a BI.  The nature of the metallic states depends on the local
features of the band structure like the presence of particle or hole
pockets where the chemical potential crosses the Fermi level.

\section{\label{sec:micro}Microscopic considerations: Derivation of a
  \jhalftitle{} model from a multi-orbital $\boldmath{t_{2g}\text{}}$
  model in the strong spin-orbit coupling limit}

Having derived the general symmetry-allowed tight-binding model in
Sec. \ref{sec:tb}, here we explore a microscopic multi-orbital
$t_{2g}$ tight-binding model with SOC on the ideal hyperhoneycomb
lattice within the Slater-Koster approximation.\cite{PhysRev.94.1498}
As noted earlier, unlike the generic tight-binding model considered in
the previous sections where oxygen and iridium distortions are
encapsulated in quantitative changes in the \jhalf{} hopping
amplitudes, here we specialize in the case where both the iridium and
oxygen ions are in their ideal positions.  This implies that each
iridium ion is surrounded by a perfect oxygen octahedron, all 1NN
bonds are of equal length, and the Ir-Ir-Ir and Ir-O-Ir bond angles
are $120^{\circ}$ and $90^{\circ}$ respectively.  With these
assumptions and in the limit of large SOC, we will show the connection
between the microscopic tight-binding model and the generic
symmetry-allowed tight-binding model presented in
Sec. \ref{sec:tb}---particularly the relations between the microscopic
Slater-Koster parameters and the hopping parameters introduced
earlier.  The results of this section will provide us with valuable
insights in the understanding of the DFT results in the next section.

In the ideal hyperhoneycomb lattice, each Ir ion resides in a perfect
octahedral cage of oxygen ions.  The resulting crystal field causes
the Ir $d$-orbitals to split into the lower energy $t_{2g}$ orbitals
and the higher energy $e_{g}$ orbitals with energy difference on the
order of a few electron-volts. When SOC and hopping amplitudes are
much smaller than the crystal field energy splitting, the $e_g$
orbitals can be projected since the 5 electrons at each Ir$^{4+}$ site
will mostly contain $t_{2g}$ character.  The atomic SOC, when
projected on the $t_{2g}$ orbitals, have the following form
\begin{equation}
  H_{\text{SOC}}=-\sum_{i} \lambda \vec{L}_{i}\cdot\vec{S}_{i},
\end{equation}
where $\vec L_i$ transforms as an angular momentum one operator (with
the three $L_z$ components being linear combinations of the three
$t_{2g}$ orbitals\cite{PhysRev.171.466}), $\vec S_i$ is the spin of a
single electron occupying the $t_{2g}$ orbitals, and $\lambda(\sim 500
meV)$ is the strength of the atomic SOC.  Due to the negative
sign\cite{PhysRev.171.466}, the \jhalf{} orbitals are higher in energy
than the \jthreehalf{} orbitals.

We consider two types of hopping amplitudes between 1NN iridium ions
within the Slater-Koster approximation: the direct overlap between
adjacent Ir $t_{2g}$ orbitals and the indirect hopping mediated by the
two shared oxygen ions in the edge-shared oxygen octahedra
configuration.  The resulting tight-binding model in the $t_{2g}$
basis can be written as
\begin{equation}
  H_{t_{2g}}=\sum_{\langle i j \rangle} d^{\dagger}_{i}\left[  h^{\text{direct}}_{ij}(t_{\sigma},t_{\pi},t_{\delta})
    + h^{\text{indirect}}_{ij}(t_{\text{oxy}}) \right]d_{j},
    \label{eq:H_t2g}
\end{equation}
where
$d^{\dagger}=\left(d^{\dagger}_{yz},d^{\dagger}_{xz},d^{\dagger}_{xy}\right)$
are the creation operators in the $t_{2g}$ basis.  The direct hopping
matrix $h^{\text{direct}}_{ij}$ is parameterized by Slater-Koster
parameters $t_{\sigma}$, $t_{\pi}$, and $t_{\delta}$ representing
$\sigma$, $\pi$, and $\delta$ hopping amplitudes between adjacent
$t_{2g}$ orbitals respectively.  The indirect hopping matrix
$h^{\text{indirect}}_{ij}$ is parameterized by
$t_{\text{oxy}}=|t_{pd\pi}|^2/\Delta$, where $t_{pd\pi}$ is the $\pi$
hopping between iridium $d$-orbitals and oxygen $p$-orbitals and
$\Delta$ is the energy difference between those two sets of
orbitals. The detailed form of the hopping matrices is outlined in
Appendix \ref{app:hopping}.

In the large SOC limit, the bands arising from the \jhalf{} and the
\jthreehalf{} orbitals are expected to separate.  In the
$\lambda\rightarrow\infty$ limit, an effective tight-binding model
involving only the \jhalf{} degrees of freedom can be obtained by
lowest order perturbation theory: projection of the $t_{2g}$ bands
into the \jhalf{} manifold
\begin{align}
  H_{\text{eff}}^{(1)}=\mathcal{P}H_{t_{2g}}\mathcal{P},
\end{align}
where $\mathcal{P}$ is the projector for the \jhalf{} manifold.  This
projection yields a 1NN \jhalf{} model with

\begin{align}
t^{1\text{NN}}_{xy}=t^{1\text{NN}}_z=\left(3 t_{\sigma} + 4 t_{\pi} + 5 t_{\delta}\right)/6;~~~ |v^{1\text{NN}}_{z}|=0.
\end{align}  
As discussed in Sec. \ref{subsec:tb1NN}, this effective Hamiltonian
is particle-hole symmetric and can only host a metallic phase with a
line-node Fermi surface.  In addition, the model is manifestly
SU$(2)$-invariant despite the presence of SOC.  Lastly,
oxygen-mediated hopping does not contribute at this order.  This is
because the amplitudes from the two oxygen-mediated hopping paths
cancel exactly under projection into the \jhalf{} manifold when
Ir-O-Ir bond angles are $90^{\circ}$ .  Clearly, the above model does
not represent the general structure and the next order correction
arising from finite values of $\lambda$ must be considered to better
describe the band structure obtained in the original $t_{2g}$ model.

Including the second order term in perturbation theory, the effective
Hamiltonian can be written as
\begin{align}
  H_{\text{eff}} &= H_{\text{eff}}^{(1)} + H_{\text{eff}}^{(2)}
  +\mathcal{O}\left(\frac{H_{t_{2g}}^3}{(3\lambda/2)^2} \right),
  \label{eq:H_eff}
\end{align}
with
\begin{align}
  H^{(2)}_{\text{eff}} &= \left(3\lambda/2\right)^{-1}\mathcal{P} H_{t_{2g}}
  \mathcal{Q} H_{t_{2g}} \mathcal{P},
\end{align}
where $\mathcal{Q}$ is the projector for the \jthreehalf{} manifold.
In addition to 1NN hopping generated from $H_{\text{eff}}^{(1)}$, the
second order term, $H^{(2)}_{\text{eff}}$, now generates 2NN hopping
amplitudes via virtual hopping to \jthreehalf{} orbitals at
intermediate Ir sites.  The generated 2NN hopping amplitudes takes the
form of those considered in Sec. \ref{subsec:tb2NN}, hence, we can
relate the Slater-Koster parameters used in this section with those
used in the generic symmetry-allowed tight-binding model.  The
relations, including the contribution from $H_{\text{eff}}^{(1)}$, are
given by
\begin{align}
  |v_{z}|&=0, \nonumber \\
  t_{xy}&=t_{z}=\left(3 t_{\sigma} + 4 t_{\pi} + 5 t_{\delta}\right)/6, \nonumber \\
  t_{2\text{NN}}&=-\left(3 t_{\sigma} - 2 t_{\pi} - t_{\delta}  \right)^2/\left(108 \lambda\right), \nonumber\\
  v_{\text{KM}}&=\left(t_{\pi} - t_{\delta} - 2
      t_{\text{oxy}}\right)
    \left(3 t_{\sigma} - 3 t_{\pi} + 2 t_{\text{oxy}} \right)/(9\sqrt{3} \lambda),\nonumber \\
  v_{\parallel}&=\left(t_{\pi} - t_{\delta} - 2
      t_{\text{oxy}}\right) \left(3 t_{\sigma} - 3 t_{\delta} - 4 t_{\text{oxy}}
    \right)/(9\sqrt{6} \lambda).
    \label{eq:t2gtojeff}
\end{align}
We note that by assuming an ideal hyperhoneycomb lattice structure
together with truncating the perturbation series at second order, the
1NN $xy$ and $z$ bonds have the same scalar-only hopping amplitude.
Furthermore, truncating the perturbation series at second order
implies that 2NN hopping amplitudes are only generated on 2NN bonds
with shared Ir sites.  These 2NN bonds, although not related by
symmetry, have related hopping amplitudes because of the assumed ideal
structure and the truncated perturbative series (see Appendix
\ref{app:2NN} for details).  Since the higher order terms in the
series fall off as an inverse power of the SOC coupling, we expect
that these higher order terms are small in magnitude and hence may be
negligible to the leading order.

By establishing the $t_{2g}$ tight-binding model in the Slater-Koster
approximation, we can perform a loose fit against \textit{ab initio}
calculations to obtain an estimate of these hopping amplitudes as we
show in the next section.  Furthermore, by relating the Slater-Koster
parameters with hopping amplitudes used in the generic tight-binding
model, short-ranged electronic correlation can be included
straightforwardly in the effective \jhalf{} model as we exemplify in
Sec. \ref{sec:mag}.

\section{\label{sec:fit}\textit{Ab Initio} calculations on
  \liirotitle{} in the ideal structure and connection to the
  tight-binding model}

Having discussed the details of the tight-binding models, we now
employ {\it ab initio} approaches for the ideal \liiro{} structure and
try to capture its characteristic features via the tight-binding model
introduced in the previous section. At the outset, we note that in the
absence of data determining the accurate lattice structure of
\liiro{}, we have assumed that it has ideal structure and the oxygen
octahedra are not distorted. While this structure may not be an
accurate description of the material, it gives us an idea of the
general validity of the approximations made in the two earlier
sections about the \jhalf{} nature of the bands near the Fermi
level. Also we can obtain a qualitative estimate of the various
parameters used in the previous two tight-binding models. We look for
general features that may aid the determination of the parameter
regime of the tight-binding Hamiltonian which is of interest in the
context of materials.

\begin{figure}[t!]
  \centering
  \captionsetup[subfigure]{labelformat=empty}
  \setlength\fboxsep{0pt}
  \setlength\fboxrule{0.pt}
  \subfloat[][]{
    \fbox{\begin{overpic}[scale=1,trim=-28 -15 0 -15,clip,tics=10]{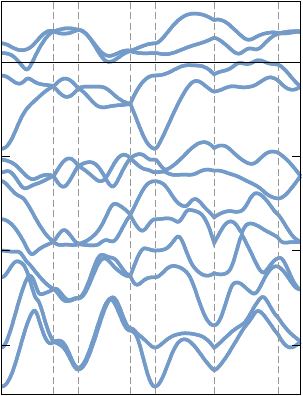}
        \put(67,135){(a)}
        \put(27,4){$\Gamma$}
        \put(40,4){Y}
        \put(48,4){T}
        \put(63,4){Z}
        \put(70,4){$\Gamma$}
        \put(86,4){X}
        \put(100,4){A$_1$}
        \put(109,4){Y}
        \put(0,66){$eV$}
        \put(15,109){0.0}
        \put(12,82){-0.5}
        \put(12,55){-1.0}
        \put(12,27){-1.5}
      \end{overpic}}
  }
  \subfloat[][]{
    \fbox{\begin{overpic}[scale=1,trim=0 -15 -19 -15,clip]{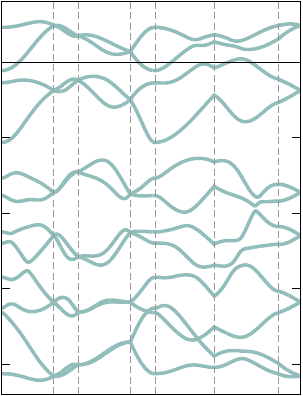}
        \put(38,135){(b)}
        \put(0,4){$\Gamma$}
        \put(12,4){Y}
        \put(20,4){T}
        \put(35,4){Z}
        \put(42,4){$\Gamma$}
        \put(59,4){X}
        \put(72,4){A$_1$}
        \put(81,4){Y}
        \put(93,109){0.0}
        \put(90,88){-0.5}
        \put(90,66){-1.0}
        \put(90,44){-1.5}
        \put(90,22){-2.0}
      \end{overpic}}
  }

  \vspace{-24pt}
  \subfloat[][]{
    \fbox{\begin{overpic}[scale=1,trim=-28 -15 0 0,clip]{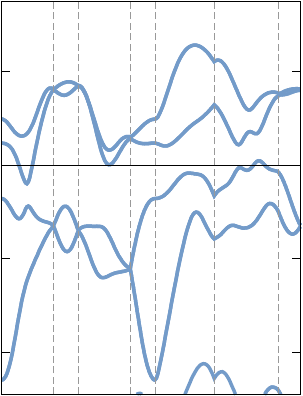}
        \put(67,5){(c)}
        \put(0,66){$eV$}
        \put(15,106){0.2}
        \put(15,79){0.0}
        \put(12,52){-0.2}
        \put(12,25){-0.4}
      \end{overpic}}
  }
  \subfloat[][]{
    \fbox{\begin{overpic}[scale=1,trim=0 -15 -19 0,clip,tics=10]{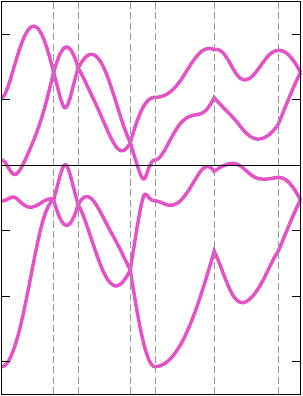}
        \put(38,5){(d)}
        \put(93,117){0.4}
        \put(93,98){0.2}
        \put(93,79){0.0}
        \put(90,61){-0.2}
        \put(90,42){-0.4}
        \put(90,23){-0.6}
      \end{overpic}}
  }
  \vspace{-10pt}
  \caption{(Color online) Electron band structures of ideal $\beta$-Li$_2$IrO$_3$ and tight-binding fit.
  (a) $t_{2g}$ bands from the DFT calculation.
  (b) $t_{2g}$ tight-binding model with the parameter in Eq. (\ref{eq:t2g_para}).
  (c) Top eight $t_{2g}$ (\jhalf{}) bands in (a).
  (d) \jhalf{} tight-binding model with the parameter in Eq. (\ref{eq:jhalf_para}). 
  In the above plots, the Fermi level is at 0 $eV$, the DFT bands are plotted in blue, the tight-binding $t_{2g}$ bands in green, and the tight-binding \jhalf{} bands in red.
  Each band is doubly degenerate due to time-reversal and inversion symmetry.
  \label{fig:DFT_bands}}
\end{figure}
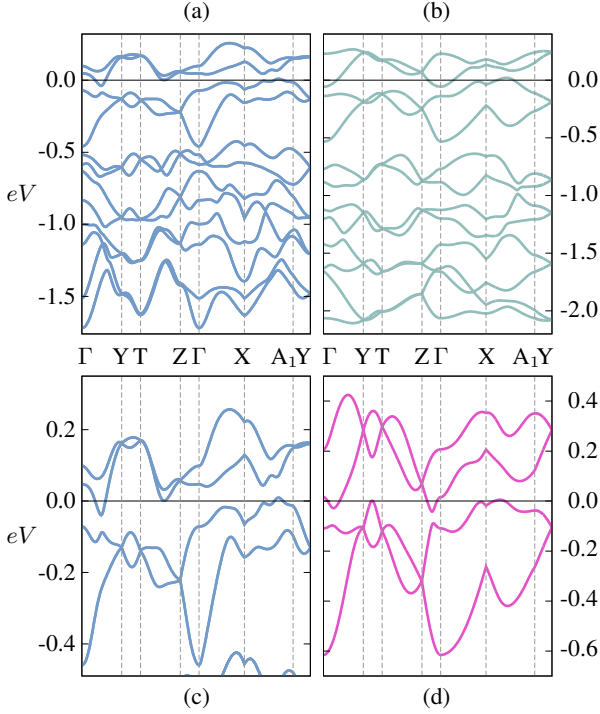
  
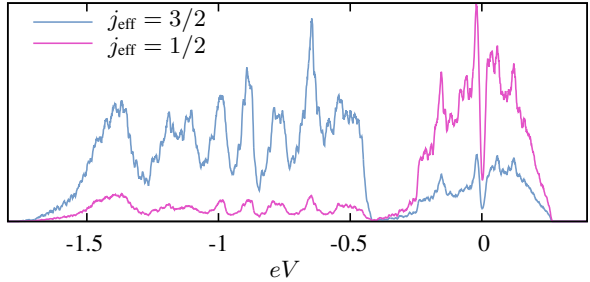
\begin{figure}
  \centering
  \setlength\fboxsep{0pt}
  \setlength\fboxrule{0.0pt}
  \fbox{\begin{overpic}[scale=1,trim=0 -5 0 0,clip]{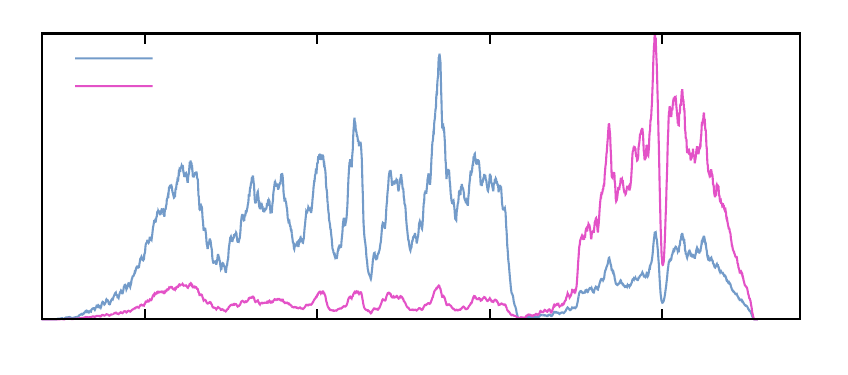}
      \put(50,85){\jhalf{}}
      \put(50,95){\jthreehalf{}}
      \put(34,10){-1.5}
      \put(88,10){-1}
      \put(134,10){-0.5}
      \put(190,10){0}
      \put(111,0){$eV$}
    \end{overpic}}
  \caption{(Color online) Density of states for the DFT band structure
    in Fig. \ref{fig:DFT_bands} (a).  The density of states is
    projected into the $j_{\textup{eff}}=1/2$ (red) and
    $j_{\textup{eff}}=3/2$ (blue) orbital sectors.}
  \label{fig:DOS}
\end{figure}

Fig. \ref{fig:DFT_bands} (a) shows twelve $t_{2g}$ bands from the DFT
calculation for the ideal \liiro{} using OpenMx\cite{boker2011openmx}
in which the linear-combination-of-pseudo-atomic-orbital formalism and
a fully-relativistic $j$-dependent pseudopotential in a non-collinear
methodology are adopted.  The Perdew-Burke-Ernzerhof generalized
gradient approximation (GGA) functional was used for the
exchange-correlation energy,\cite{perdew1996generalized} and 300Ry of
energy cutoff and the 12 $\times$ 12 $\times$ 12 Monkhorst-Pack grid
are used for the real- and the momentum-space integrations,
respectively.  Each of the twelve bands is doubly degenerate due to
time-reversal and inversion symmetries. A remarkable feature of the
bands near the Fermi level is their pronounced \jhalf{} character. As
the density of states (DOS) of the band structure shows
(Fig. \ref{fig:DOS}), the upper four bands ($\gtrsim -0.5 ~eV$) have
strong $j_{\text{eff}}=1/2$ orbital character while the bottom eight
bands ($\lesssim -0.5 ~eV$) have main contributions coming from
$j_{\text{eff}}=3/2$ orbitals. Hereafter, we call the former
$j_{\text{eff}}=1/2$ bands and the latter $j_{\text{eff}}=3/2$
bands. Another notable result is that the ideal \liiro{} structure is
in a metallic phase in the non-interacting limit. The Fermi level
crosses the $j_{\text{eff}}=1/2$ bands generating several Fermi
pockets along the lines $\Gamma$-Y, T-Z, X-A$_1$, etc. [see
Fig. \ref{fig:DFT_bands} (c)].
  
The predominant \jhalf{} character of the bands near the Fermi level
strongly supports our assumption that the low energy electronic
degrees of freedom can be adequately described by \jhalf{} orbitals.
In turn, this lends credence to our use of the \jhalf{} tight-binding
model in the previous sections when modeling Ir-based hyperhoneycomb
compounds.

Next, we fit the $t_{2g}$ model in Eq. (\ref{eq:H_t2g}) to the DFT
results by adjusting the Slater-Koster parameters. Figure
\ref{fig:DFT_bands} (b) shows the resulting electronic bands of the
$t_{2g}$ model, which has following hopping parameters:
\begin{equation}
 \begin{array}{lcrlcr}
 t_{\sigma} &=& -0.4574 ~ eV, & t_{\pi} &=& 0.6098 ~ eV,
 \\
 t_{\delta} &=& -0.0041 ~ eV, & t_{oxy} &=& 0.1155 ~ eV.
 \end{array}
 \label{eq:t2g_para}
\end{equation}
In our fitting process, we adopted $ \lambda = 0.5797 ~ eV$ from
Ref. \onlinecite{micklitz2010spin} and adjusted the other parameters,
$t_{\sigma},~t_{\pi},~t_{\delta},$ and $t_{oxy}$. The tight-binding
model reproduces two overall features found in the DFT computation:
(1) well-separated \jhalf{} (top four) and \jthreehalf{} (bottom
eight) bands and (2) a semi-metallic phase, albeit with pockets at
different positions from those found in the DFT. However, quantitative
details like correct band curvatures and energy values are not
recovered within our model, indicating that further neighbor hopping
amplitudes are required for a quantitatively better fit. According to
our Wannier function analysis within DFT,\cite{kim2014unpublished} it
is necessary to include up to fourth-nearest-neighbor hopping
amplitudes in the tight-binding model to recover the quantitative
features of the DFT band structure, thus we should take our
tight-binding fit as a ``loose'' fit that aims not to replicate exact
details but to reproduce qualitative features of the DFT
results. Among the hopping amplitudes up to the
fourth-nearest-neighbors, the 1NN hopping amplitudes have the largest
magnitudes and determine the overall behavior of the band structure
while further neighbor hopping amplitudes, which have relatively small
magnitudes, are responsible for detailed structures. This justifies
the calculations in the previous section where we have only taken the
1NN hopping amplitude in the $t_{2g}$ Hamiltonian to be non-zero.

By mapping the $t_{2g}$ model obtained from the fitting procedure to
the effective \jhalf{} model in Eq. (\ref{eq:H_eff}), we
arrive at the following values for the hopping amplitudes via the
relations given in Eq. (\ref{eq:t2gtojeff}):
\begin{equation}
 \begin{array}{lcr}
 t_{xy,z} &=& 0.1744 ~ eV,
 \\
 t_{\textup{2NN}} &=& -0.1150 ~ eV,
 \\
 v_{\textup{KM}} &=& -0.1331 ~ eV,
 \\
 v_{\parallel} &=& -0.0222 ~ eV,
 \end{array}
 \label{eq:jhalf_para}
\end{equation}
which corresponds to a point in the metallic region of
Fig. \ref{fig:2NNpd}, with $t_{\textup{2NN}}^2+v_{\parallel}^2 >
0.5^2$.  The band structure of the resulting \jhalf{} model is plotted
in Fig. \ref{fig:DFT_bands} (d) for comparison with the band
structures of the $t_{2g}$ model and DFT results.

This concludes our discussions on the electronic structure. Below we
shall investigate the effect of correlations in stabilizing magnetic
ordering in the intermediate correlation regime.

\section{\label{sec:mag}Magnetic order at intermediate coupling}

In several iridate compounds where the $5d^5$ iridium ions are
octahedrally-coordinated with oxygen ions, magnetic ordering often
occur at the iridium sites due to short-ranged electronic
correlations.\cite{witczak2013correlated} As correlations are
increased, the system changes from a paramagnetic metal to a
magnetically ordered metal which at higher correlations becomes an
insulator.  In certain instances, the magnetic ordering and the
metal-insulator transitions have been observed to occur
simultaneously.

Here we explore this scenario in the intermediate coupling regime on
the hyperhoneycomb lattice via self-consistent mean-field theory of
the \jhalf{} model. Starting with the 2NN tight-binding model outlined
in Eq. (\ref{eq:H_eff}), we include correlation effects via on-site
Hubbard repulsion
\begin{equation}
U\sum_{i} n_{i\uparrow}n_{i\downarrow} = -\frac{2U}{3}\sum_{i}\mathbf{J}_{i}\cdot\mathbf{J}_{i} + \frac{U}{2}\sum_{i} n_{i}, 
\label{eq:hubbard}
\end{equation}
where $U>0$ is the Hubbard repulsion strength, $n_{i\sigma}$ is the
number operator at site $i$ with pseudo-spin $\sigma$,
$n_{i}=\sum_{\sigma}n_{i \sigma}$, and $\mathbf{J}_{i}$ is the
\jhalf{} pseudo-spin operator. The local magnetic moment, when
projected into the \jhalf{} manifold, is proportional to the local
\jhalf{} moment, \textit{i.e.} $\mathbf{M}_i=-2\mathbf{J}_i$.  Hence,
a Hartree-Fock decoupling of the $\mathbf{J}_{i}\cdot\mathbf{J}_{i}$
term will yield a mean-field Hamiltonian that can be self-consistently
solved for the magnetic ordering of the \jhalf{} moments. In the
absence of compelling experimental motivation to choose particular
hopping amplitudes, we choose a cut which interpolates between the
purely isotropic 1NN model and the tight-bonding model whose
parameters are given by our DFT calculations in
Eq. (\ref{eq:jhalf_para}). This is done in the following way: we
choose
\begin{equation}
 \begin{array}{lcr}
   t_{xy,z} &=& 0.1744 ~ eV,
   \\
   t_{\textup{2NN}} &=& (-0.1150 ~ eV)x,
   \\
   v_{\textup{KM}} &=& (-0.1331 ~ eV)x,
   \\
   v_{\parallel} &=& (-0.0222 ~ eV)x.
 \end{array}
\end{equation}
and then vary $x$ between $0$ and $1$ to interpolate between the two
above limits. With this particular choice of hopping amplitudes, we
are able to explore the effects of correlation on both the STI phase
($x\lesssim 0.74$) and the metallic phase ($x\gtrsim 0.74$).

To perform the self-consistent mean-field calculations, we consider
four 3-component order parameters---$\langle \mathbf{J}_{i} \rangle$
with $i=1\dotsc 4$---and assume $\mathbf{q}=\mathbf{0}$
order.\footnote{We have explicitly checked that the spin model
  obtained in the strong coupling expansion of our Hubbard model
  always yielded the same $\mathbf{q}=\mathbf{0}$ magnetic order as
  our mean-field results within the parameter regime we considered.
  Moreover, this magnetic order remains the ground state for a broad
  range of parameters in the spin model as long as the 2NN spin-spin
  interactions remain moderately small.  This fact gives us reason to
  believe that $\mathbf{q}= \mathbf{0}$ ansatz in our mean-field
  calculation may be a reasonable simplification.}  The
self-consistent solution is achieved with no constraints on the
magnetic configuration such that all $\mathbf{q}=\mathbf{0}$ ordering
can be sampled in principle.

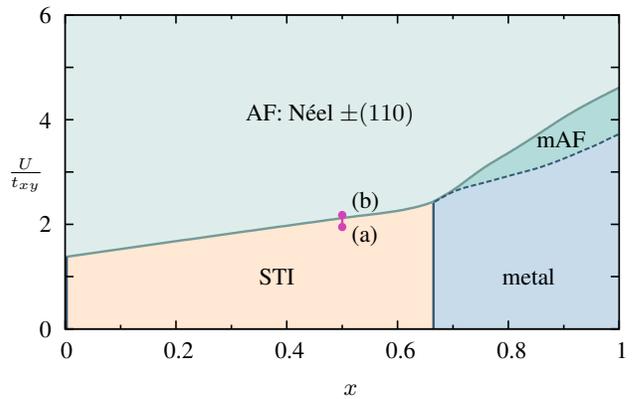
\begin{figure}
  \centering
  \setlength\fboxsep{0pt}
  \setlength\fboxrule{0.0pt}
  \fbox{\begin{overpic}[scale=1,trim=0 -10 0 0,clip]{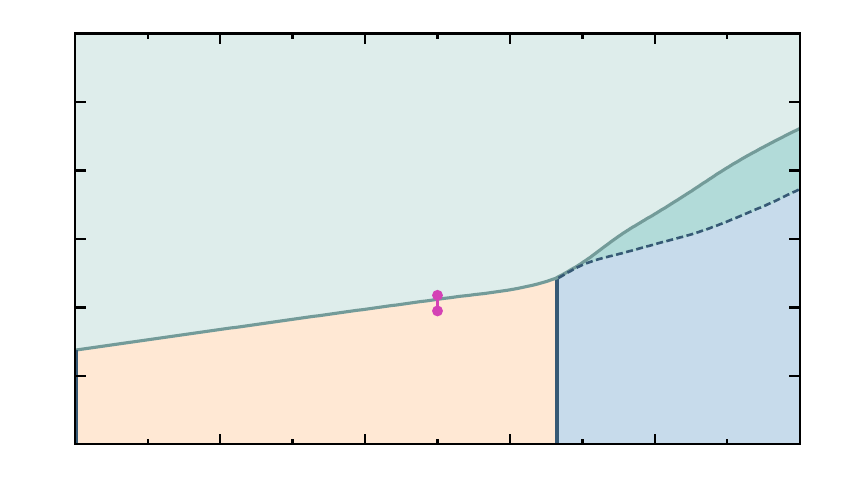}
      \put(0,83){$\frac{U}{t_{xy}}$}
      \put(127,1){$x$}
      \put(90,105){AF: N\'eel $\pm(110)$}
      \put(95,43){STI}
      \put(187,43){metal}
      \put(200,95){mAF}
      \put(130,72){(b)}
      \put(130,59){(a)}
      \put(20,15){0}
      \put(59,15){0.2}
      \put(101,15){0.4}
      \put(143,15){0.6}
      \put(185,15){0.8}
      \put(230,15){1}
      \put(12,23){0}
      \put(12,63){2}
      \put(12,102){4}
      \put(12,142){6}
    \end{overpic}}
  \caption{(Color online) Phase diagram at finite Hubbard repulsion
    $U$ as a function of 2NN hopping strength $x$ (see text for
    description of 2NN hopping used).  The local magnetic moments of
    the N\'eel state (AF) and the metallic magnetically ordered phase
    (mAF) point in the $\pm(110)$ direction.  The solid line between
    the strong topological insulator (STI) and the AF state indicates
    a second-order transition while the dashed line between the metal
    and the mAF phases indicates a first-order transition.  The
    transition from the mAF phase to the AF state is continuous.  The
    $x=0$ line describes a purely 1NN model where the paramagnetic
    ground state has a line-node Fermi surface, as indicated by the
    blue line.  The two points, (a) and (b), indicate where the slab
    configurations were computed in Fig. \ref{fig:slab}}
  \label{fig:hf_pd}
\end{figure}

In Fig. \ref{fig:hf_pd}, we present the finite $U$ phase diagram as a
function of 2NN hopping amplitudes. Upon increasing $U/t_{xy}\gtrsim
1.4-3.4$, the time-reversal symmetric phases undergo phase transitions
to an antiferromagnetic, N\'eel ordered phase with magnetic moments
pinned along the $+\hat{x}+\hat{y}$ direction (light green).  From the
STI (orange), the phase transition is of second order.  On the other
hand, starting with the metallic phase (blue), a first order
transition is observed.  This first order transition initially brings
the system into a magnetically ordered metallic phase (mAF, dark
green), then upon further increase in $U$, the system acquires a
finite excitation gap and becomes insulating. This metal-insulator
transition is continuous in the magnetic order parameters.  We also
note that at $x=0$, the model reduces to a purely 1NN model and the
paramagnetic phase is metallic with a line-node Fermi surface as
outlined in Sec. \ref{subsec:tb1NN}.  This phase is indicated as the
vertical blue line at $x=0$ running along the $U/t_{xy}$ axis.

Although this magnetic order breaks inversion symmetry ($\mathcal{P}$)
of the lattice, it preserves inversion followed by time reversal
symmetry ($\mathcal{P}\cdot \Theta$).  Since $\mathcal{P}(\mathbf{k})$
commutes with $\Theta(\mathbf{k})$ for all $\mathbf{k}$ in the
Brillouin zone of the hyperhoneycomb lattice, a pseudo-Kramers
degeneracy is present at all momenta.  Instead of the usual Kramers
degeneracy where $\Theta$ protects the degeneracy and together with
$\mathcal{P}$ ensure at least doubly degenerate bands, these
pseudo-Kramers bands are protected by the combined operation
$\mathcal{P}\cdot \Theta$.

\begin{figure}
  \centering
  \setlength\fboxsep{0pt}
  \setlength\fboxrule{0.pt}
  \subfloat[][]{
    \fbox{\begin{overpic}[scale=1,trim=-28 -10 3 -10,clip]{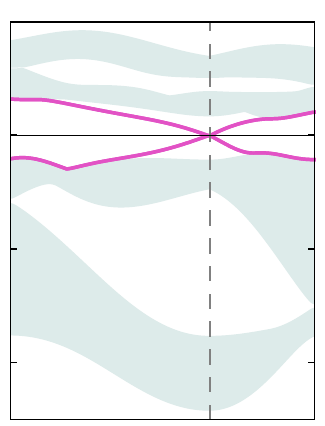}
        \put(0,71){$eV$}
        \put(17,124){0.2}
        \put(17,91){0.0}
        \put(14,59){-0.2}
        \put(14,26){-0.4}
        \put(58,132){$U<U_{\text{crit}}$}
        \put(86,00){$\bar{\Gamma}$}
        \put(30,00){$\bar{\text{X}}$}
        \put(114,00){$\bar{\text{Y}}$}
      \end{overpic}}
  }\hspace{0pt}
  \subfloat[][]{
    \fbox{\begin{overpic}[scale=1,trim=3 -10 -15 -10,clip]{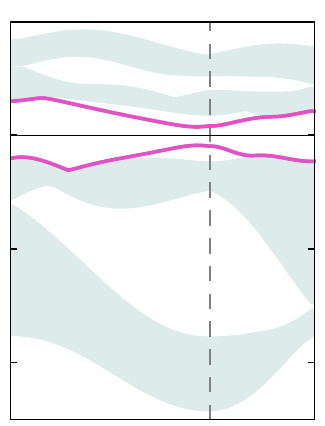}
        \put(94,124){0.2}
        \put(94,91){0.0}
        \put(91,59){-0.2}
        \put(91,26){-0.4}
        \put(28,132){$U\gtrsim U_{\text{crit}}$}
        \put(55,0){$\bar{\Gamma}$}
        \put(0,0){$\bar{\text{X}}$}
        \put(83,0){$\bar{\text{Y}}$}
      \end{overpic}}
  }
  \caption{\label{fig:slab}(Color online) Gapping out the gapless
    surface states.  For illustration, we present surface states from
    a slab configuration.  Purple bands are surface states while the
    shaded regions are the projected bulk bands.  We have set 2NN
    hopping amplitudes to be $x=0.5$ (see main text for definition and
    Fig. \ref{fig:hf_pd} for reference).  The system undergoes a
    second order transition from a strong topological insulator (STI)
    to a magnetically ordered phase (AF) at $U_{\text{crit}}\simeq
    2.12 t_{xy}$.  In (a), $U<U_{\text{crit}}$, the ground state is a
    STI and a surface Dirac cone is seen at $\bar{\Gamma}$.  In (b),
    $U$ is slightly above $U_{\text{crit}}$, the ground state is the
    AF phase, and the surface Dirac cone develops a finite gap.}
\end{figure}
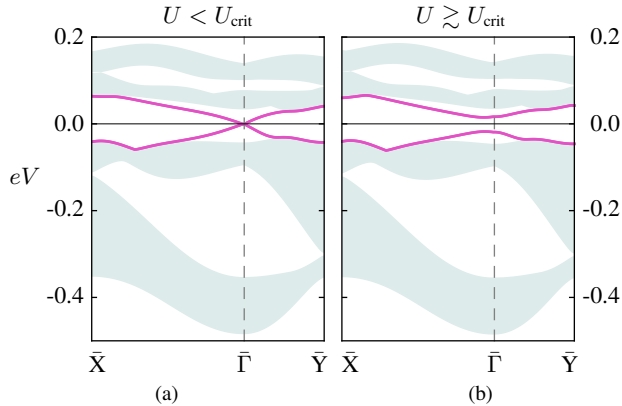

With $\mathcal{P}\cdot \Theta$ playing the role of a preserved
anti-unitary symmetry, the magnetically ordered state may harbour
non-trivial topology in the spirit of
Ref. \onlinecite{mong2010antiferromagnetic,fang2013topological,liu2013antiferromagnetic,zhang2014topological}.
However, in the present case, we find that the magnetic phase has a
trivial band structure and there are no gapless surface states arising
from non-trivial band topology. We show this in Fig. \ref{fig:slab},
where band structure calculations in a slab configuration are
presented with 2NN hopping amplitudes set to $x=0.5$. Here, the
$\mathbf{a}_1$ direction (see Fig. \ref{fig:lattice}) has finite
spatial extent.\footnote{Similar results from slab calculations with
  finite spatial extent in the $\mathbf{a}_{2}$ and $\mathbf{a}_3$
  have also been obtained, but not shown.}  While in the STI phase,
the surface Dirac cone at the $\bar{\Gamma}$-point can be seen.
However, upon increasing $U$ through the second order transition, the
surface Dirac cone becomes fully gapped (nearby momentum points were
also checked to assure that the gap was fully developed).

\section{\label{sec:discuss}Discussion and Outlook}

In this work, we have investigated the weak and intermediate coupling
regime of iridium-based compounds on the hyperhoneycomb lattice.
Using a combination of symmetry arguments and a more microscopic
calculation based on the Slater-Koster approximation, we have
determined the low energy electronic structure for the \jhalf{}
orbitals by explicitly constructing a tight-binding model. Such
orbitals are expected to dominate the low energy physics of Ir due to
strong SOC. Our DFT calculations for \liiro{} supports this
expectation by showing that the bands near the Fermi level indeed have
a \jhalf{} character. The tight-binding model admits both trivial as
well as topological band insulators along with a metallic phase.

We study the effects of Hubbard-type electronic correlations on the
above band structure, particularly in the stabilization of magnetic
order. Restricting ourselves to $\mathbf{q}=\mathbf{0}$ magnetic
orders, we perform Hartree-Fock calculations and determine the mean
field phase diagram. Our calculations admit a direct continuous phase
transition between the STI and the N\'eel ordered magnetic insulator
as we turn on correlations. The magnetic insulator breaks both
time-reversal symmetry and inversion symmetry independently, but it
preserves the product of the two. Though not realized within the
current parameter regime, this raises a possibility of a concrete
microscopic model on the hyperhoneycomb lattice where such an
anti-unitary symmetry (product of time reversal an inversion) may
stabilize a non-trivial topological band structure for the electrons
in the presence of interactions. We also note that while our
calculations have been directly motivated by the recent discovery of
\liiro, the results presented here are not necessarily restricted to
this particular material.

It is instructive to consider the range of applicability of our
current work and contrast it with previous approaches. Previous works
on Ir-based compounds on the hyperhoneycomb lattice studied the
Heisenberg-Kitaev spin Hamiltonian, which may be applicable in the
strong-coupling
regime.\cite{PhysRevB.89.045117,PhysRevB.89.014424,kimchi2013,hermanns2014quantum}
In these works, the correlation effects of Hund's coupling is
paramount to the presence of the Kitaev interaction from a microscopic
perspective.\cite{nussinov2013compass,PhysRevLett.105.027204,PhysRevLett.102.017205}
Implicit in the derivation is the assumption of $U>(J_H,\lambda) \gg
t$ where $J_H$ is Hund's coupling and $t$ is the typical hopping
amplitude.  In contrast, in Sec. \ref{sec:mag} of the present work, we
considered the intermediate-coupling limit where $\lambda \approx
U\approx t \gg J_H$ such that the effects of \jthreehalf{} states can
be treated perturbatively and that Hund's coupling can be ignored.  In
addition, other theoretical approaches have been explored in other
iridate
compounds.\cite{PhysRevB.84.100406,PhysRevB.88.035107,rau2013generic}
Indeed, these ideas could stimulate interesting future research
directions in the theoretical study of \liiro{}.

Although the definitive structure of \liiro{} is presently not known,
the \jhalf{} orbitals may still be the lowest energy degree of freedom
under sufficiently small distortions and non-octahedral crystal field
effects.\cite{gretarsson2013crystal,sohn2013mixing} If
further-neighbor hopping amplitudes are negligible, the tight-binding
calculations presented in Sec. \ref{sec:tb} will be applicable, though
the parameters in the more microscopic calculations may be affected.
If distortions are small, however, the results of our microscopic
calculations may have captured the essential qualitative features
present in the electronic structure \liiro{} and other,
yet-to-be-discovered, iso-structural iridate compounds.  Furthermore,
the magnetic order that may be present in these compounds may be
well-described by our mean-field calculations if the material lies
within the intermediate coupling regime.  In this regard, we believe
that our results serve as a valuable starting point in the description
of these fascinating compounds.

\acknowledgements We thank S.-B. Lee and V. Vijay Shankar for
discussions. This research was supported by the NSERC, CIFAR, and
Centre for Quantum Materials at the University of Toronto.  H.-S. Kim
was supported by Basic Science Research Program through the National
Research Foundation of Korea(NRF) funded by the Ministry of
Education(Grant No. 2013R1A6A3A01064947).  H. Jin is supported by the
Research Center Program of the Institute for Basic Science in Korea.

\bibliography{hhc_tb}

\appendix

\section{\label{app:lattice}Hyperhoneycomb lattice}
The space group of the hyperhoneycomb lattice is $Fddd$.  The Ir ions
occupy the Wyckoff position $16e$, which possesses a site symmetry of
$C_2$.  These $C_2$ axes coincide with the $z$-bonds and can be used
to relate $x$- and $y$-bonds.  In addition, the positions of the Ir
ions implies that the bond center of the $z$-bonds are located at the
Wyckoff position $8a$, which possesses the site symmetry group $D_2$
with three $C_2$ axes, while the bond center of the $x$- and $y$-bonds
are located at the Wyckoff position $16c$, which are inversion centers
of the lattice.  By assigning each Ir site with orbitals that
transform like a spinor (\textit{e.g.} \jhalf{} orbitals), the
bond-center symmetry operations outlined above constraints the 1NN
tight-binding model to take the form outlined in
Sec. \ref{subsec:tb1NN}.  The 2NN bonds, on the other hand, are
less constrained by symmetry and will be discussed separately in
Appendix \ref{app:2NN}.

\section{\label{app:2NN}Second-nearest-neighbors}
In the ideal hyperhoneycomb, there are four symmetry-inequivalent sets
of 2NN bonds of equal bond length.  As mentioned in
Sec. \ref{subsec:tb2NN}, these bonds can first be classified as those
that can be connected by traversing through only one intermediate Ir
site (type 1; there are twelve such bonds), and those that cannot
(type 2; there are eight such bonds).  Furthermore, type 1 can be split
into two sets: bonds of type 1a connect different sublattices of the
same parity (\textit{i.e.} 1 with 3, 2 with 4; there are eight of
these) while bonds of type 1b connect same sublattices (\textit{i.e.}
1 with 1, etc.; there are four of these).  Type 2 bonds are also split
into two additional subclasses: bonds of type 2a connect sublattices
of different parity (\text{i.e.} 1 with 4, 2 with 3; there are four of
these) while bonds of type 2b connect same sublattices (\textit{i.e.}
1 with 1, etc.; there are four of these).  These four types of 2NN
bonds are inequivalent in that no symmetry operation can relate bonds
of different types.

The symmetry operations at the bond centers constrains the terms that
can appear in the \jhalf{} tight-binding model.  For type 1a, the bond
center does not possess any site symmetries, hence the vector hopping
along this bond can have three independent components.  For type 1b,
the bond center has a $C_2$ symmetry, hence we can choose to
parametrize the two independent components of the vector hopping with
$v_{\text{KM}}$ and $v_{\parallel}$ as outlined in
Sec. \ref{subsec:tb2NN}.  From symmetry analysis, type 1a and type 1b
bonds are unrelated.  However, in the ideal hyperhoneycomb, the
\textit{local} oxygen and lithium environments of these two types of
bonds are identical (local as defined by nearest-neighbors to the Ir
sites of the bond).  Treating non-local ions as negligible symmetry
breaking terms, we can use the mirror operation that relates these two
bond environments to relate the vector hopping of type 1a to that of
type 1b.  This simplification was used in parameterizing the 2NN bonds
in Sec. \ref{subsec:tb2NN} and is manifest in Sec. \ref{sec:micro}.

For bonds of type 2a, there exists a $C_{2}$ symmetry operation which
reduces the vector hopping amplitude to two components.  For bonds of
type 2b, the bond center is an inversion center and hence only scalar
hopping is allowed.  Bonds of type 2 were not included in our model:
this is motivated by our microscopic derivation in
Sec. \ref{sec:micro}.

\section{\label{app:ph}Particle-hole symmetry}

We consider particle-hole symmetry transformations of the following
form
\begin{align}
c_{i\sigma}\rightarrow c^\dagger_{i\sigma}~~{\rm for}~~i\in 1,3\nonumber\\
c_{i\sigma}\rightarrow -c^\dagger_{i\sigma}~~{\rm for}~~i\in 2,4.
\end{align}
We first consider scalar hopping terms.  The 1NN scalar hopping terms
in Eq. (\ref{eq:tb1nnxy}) and Eq. (\ref{eq:tb1nnz}) transform as
\begin{align}
t_{\alpha} \left(c_{i\sigma}^\dagger c_{j\sigma}+h.c.\right)\rightarrow t_{\alpha} \left(c_{i\sigma}^\dagger c_{j\sigma} + h.c.\right)
\end{align}
where $\alpha$ stands for the $x,y$ or $z$ bonds.  This shows that 1NN
scalar hopping terms are invariant.  The 2NN scalar hoping
\begin{align}
t_{\rm 2NN}\left(c_{i\sigma}^\dagger c_{j\sigma}+h.c.\right)\rightarrow -t_{\rm 2NN}\left(c_{i\sigma}^\dagger c_{j\sigma} + h.c.\right)
\end{align}
is not invariant. Hence, the $t_{\rm 2NN}$ term breaks particle-hole
symmetry.

Vector hopping terms takes the form
\begin{align}
i\left(c^\dagger_{i}{\bf v}_{ij}\cdot\mathbf{\sigma}c_{j}+c^\dagger_j{\bf v}_{ji}\cdot\mathbf{\sigma}c_i\right)
\label{eq_vechop}
\end{align}
with ${\bf v}_{ji}=-{\bf v}_{ij}$ an ${\bf v}_{ij}\in \mathbb{R}$. For the
above particle-hole transformation followed by a global U(1) spin
rotation,
\begin{align}
c_{i}\rightarrow e^{-i\frac{\pi}{4}\sigma^z}c_i,
\end{align}
we find that the 1NN vector hopping amplitude as well as the Kane-Mele
type of 2NN vector hopping amplitude in Eq. (\ref{eq:km}) are
invariant under the combined transformation. However the second
contribution to the 2NN vector hopping given by Eq. (\ref{eq:par}) is
not.

Since the U(1) rotation is global, it does not affect the invariance
of the scalar hopping amplitudes discussed above.

\section{\label{app:hopping}Hopping amplitudes in the $t_{2g}$ model}
The hopping amplitudes between 1NN can be broken up into contributions
from the direct overlap of adjacent Ir $t_{2g}$ orbitals and
oxygen-mediated hopping.  On the $z$-bond, the former, as parametrized
by Slater-Koster parameters, is given by
\begin{equation}
  h^{\text{direct}}_{z} = 
  \begin{pmatrix}
    (t_{\pi}+t_{\delta})/2 & (t_{\pi}-t_{\delta})/2 & 0 \\
    (t_{\pi}-t_{\delta})/2 & (t_{\pi}+t_{\delta})/2 & 0 \\
    0 & 0 & (3t_{\sigma} +t_{\delta})/4
  \end{pmatrix},
\end{equation}
and the latter is given by
\begin{equation}
  h^{\text{indirect}}_{z} = 
  \begin{pmatrix}
    0 & -t_{\text{oxy}} & 0 \\
    -t_{\text{oxy}} & 0 & 0 \\
    0 & 0 & 0
  \end{pmatrix},
\end{equation}
where the basis used is given by
$d^{\dagger}=\left(d^{\dagger}_{yz},d^{\dagger}_{xz},d^{\dagger}_{xy}\right)$.
The Slater-Koster parameters $t_{\sigma}$, $t_{\pi}$, and $t_{\delta}$
represent $\sigma$, $\pi$, and $\delta$ hopping amplitudes between
adjacent $t_{2g}$ orbitals respectively.  The oxygen-mediated hopping
is given by $t_{\text{oxy}}=|t_{pd\pi}|^2/\Delta$, where $t_{pd\pi}$
is the $\pi$ hopping between iridium $d$-orbitals and oxygen
$p$-orbitals and $\Delta$ is the energy difference between those two
sets of orbitals.  In the ideal hyperhoneycomb lattice, the
\textit{local} environment surrounding the $z$ and $x/y$ bonds are
related by $C_{3}$ rotations about the $(111)$-direction, hence the
hopping amplitudes on the $x/y$ bonds can be obtained by rotating the
above hopping matrices in the appropriate manner.

\end{document}